\documentclass[twocolumn]{aastex63}

\usepackage{gensymb}
\usepackage{amsmath}

\usepackage{threeparttable}

\shorttitle{The Guitar's Filament}
\shortauthors{de Vries et al.}

\begin{document}

\title{A Quarter Century of Guitar Nebula/Filament Evolution}

\correspondingauthor{Martijn de Vries}
\email{mndvries@stanford.edu}

\author{Martijn de Vries}
\affiliation{Department of Physics/KIPAC, 
Stanford University, Stanford, CA 94305-4060, USA}
\author{Roger W. Romani}
\affiliation{Department of Physics/KIPAC, 
Stanford University, Stanford, CA 94305-4060, USA}
\author{Oleg Kargaltsev}
\affiliation{Department of Physics, George Washington University, Washington, DC 20052, USA}
\author{George Pavlov}
\affiliation{Department of Astronomy \& Astrophysics, Pennsylvania State University, University Park, PA 16802, USA}
\author{Bettina Posselt}
\affiliation{Department of Astronomy \& Astrophysics, Pennsylvania State University, University Park, PA 16802, USA}
\affiliation{Department of Physics, University of Oxford, OX1 3PU Oxford, UK}
\author{Patrick Slane}
\affiliation{Harvard-Smithsonian Center for Astrophysics, 60 Garden Street, Cambridge, MA 02138, USA}
\author{Niccolo' Bucciantini}
\affiliation{INAF - Osservatorio Astrofisico di Arcetri, L.go Fermi 5, 50125, Firenze, Italy}
\author{C.-Y. Ng}
\affiliation{Department of Physics, The University of Hong Kong, Pokfulam Road, Hong Kong}
\author{Noel Klingler}
\affiliation{Department of Astronomy and Astrophysics, The Pennsylvania State University, 525 Davey Laboratory, University Park, PA 16802, USA}

\begin{abstract}
We have collected a new deep {\it Chandra X-ray Observatory} ({\it CXO}) exposure of PSR B2224+65 and the `Guitar Nebula', mapping the complex X-ray structure. This is accompanied by a new {\it HST} H$\alpha$ image of the head of the Guitar. Comparing the {\it HST} and {\it CXO} structures in 4 epochs over 25 years, we constrain the evolution of the TeV particles that light up the filament. Cross-field diffusion appears to be enhanced, likely by the injected particles, behind the filament's sharp leading edge, explaining the filament width and its evolving surface brightness profile.
\bigskip
\end{abstract}

\keywords{stars: neutron — pulsars: individual (PSR B2224+65)}

\bigskip

\section{Introduction} 
\label{sec:intro}

A handful of fast-moving pulsars have been seen to have narrow X-ray structures (`filaments') extending from the pulsar point source at large angle to the proper motion axis, itself often marked by a Pulsar Wind Nebula (PWN) trail. The first discovered, and arguably most spectacular, is that associated with PSR B2224+65. This pulsar has a very large proper motion $\mu = 194.1 \pm 0.2$\,mas/y, which at its VLBI-measured $0.83^{+0.17}_{-0.10}$\,kpc parallax distance \citep{Deller2019} implies a highly supersonic $v_{\rm PSR}=765$\,km/s. This leads to the formation of its remarkable H$\alpha$ bow shock, `The Guitar Nebula' \citep{CRL1993}, which has a very small $\sim 0.1^{\prime\prime}$ standoff angle, and a long trail forming the neck and body of the Guitar. Projecting from near the pulsar at $\sim 115^\circ$ to its proper motion is an X-ray filament, with a variable width of $\sim 20^{\prime\prime}$ and length $\sim 2.5^\prime$. It has a sharp leading edge, in the direction of the pulsar motion with surface brightness trailing off behind. 

In the picture sketched by \cite{Bandiera2008} and explored numerically by \citet{2019MNRAS.485.2041B} and \citet{2019MNRAS.490.3608O}, pulsar filaments are created by multi-TeV pulsar $e^\pm$ leaking out near the bow shock apex and escaping to external ISM field lines. 
Electrons whose gyroradius $r_c$ approaches or exceeds the standoff distance $r_0= [{\dot E}/(4\pi \mu m_p n_0 c v_{\rm PSR}^2)]^{1/2}$ may escape to the filaments. Small $r_0$ requires some combination of low pulsar $\dot{E}$, high velocity $v_{\rm PSR}$, and high ambient ISM density $n_0$. The presence of an H$\alpha$ bow shock in B2224 (implying high ISM density), as well as its large transverse velocity seem to support this picture.

In \cite{deVries2022}, we argued that the morphology of the filament associated with PSR J2030+4415 can be explained by a variable particle injection rate: the H$\alpha$ morphology suggests a temporary decrease in the standoff distance, subsequently enabling enhanced particle injection into the ISM over a period of approximately a decade. In order to connect the properties of the X-ray filament to the level of particle injection, multi-epoch H$\alpha$ bow shock observations are crucial as they allow for direct measurement of $r_0$ and provide important clues to fluctuations in $\dot{E}$ or $n_0$ at earlier times. The Guitar, which has been observed several times by both \textit{Chandra} and \textit{HST} over a period of more than 25 years, thus provides us with a unique opportunity to witness the time evolution of a pulsar filament.

The filament of B2224 has been the subject of three previous \textit{Chandra} ACIS exposures in 2000, 2006 and 2012. Since there were significant changes between these epochs, we have collected a new, deeper ACIS exposure to provide a fiducial structure and context for the earlier observations. The evolution and X-ray spectral parameters give information about the multi-TeV $e^\pm$ injection which lights up the filament. Concurrent with the new epoch of X-ray observations, we have obtained a new \textit{HST} ACS/WFC H$\alpha$ image of the apex of the nebula. 

\begin{figure*}[b]
\centering
\vskip -1cm
\includegraphics[width=\textwidth]{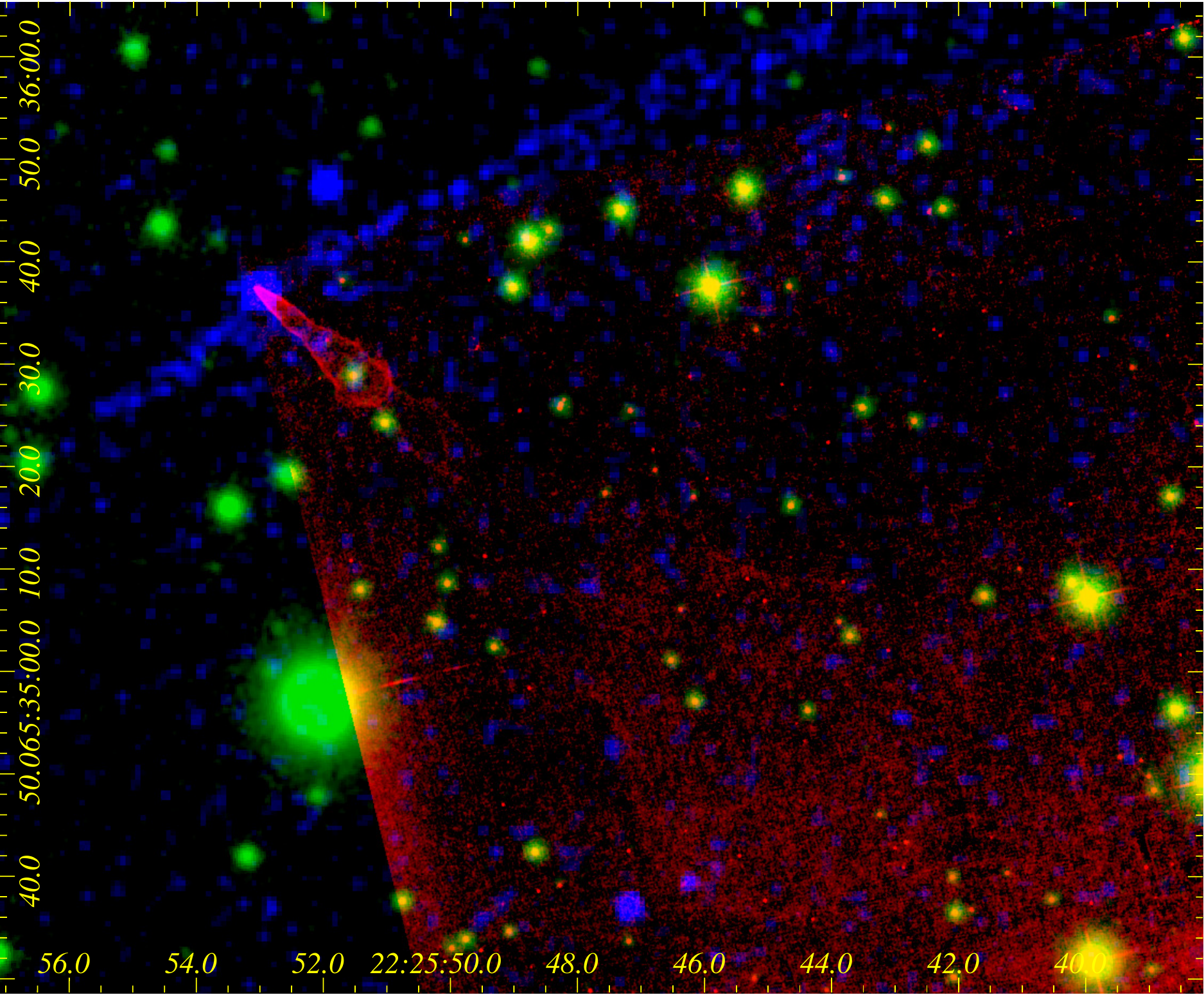}\\
\caption{Overview of the Guitar/filament complex in the most recent \textit{HST} and \textit{Chandra} epochs. Red: 2020 {\it HST} ACS/WFC H$\alpha$, green: PanSTARRS2 $r$, blue: 2021 {\it CXO} 1-5\,keV X-rays. The pulsar point source lies at the tip of the H$\alpha$ nebula. Filament X-ray emission extends primarily to the right (NW). This has a sharp leading edge and extension behind. The Guitar `body' shows faintly in the H$\alpha$ in the lower half of the image.}
\label{fig:largescale}
\end{figure*}

\begin{figure*}[h]
\centering
\vskip -3cm
\includegraphics[width=7 in]{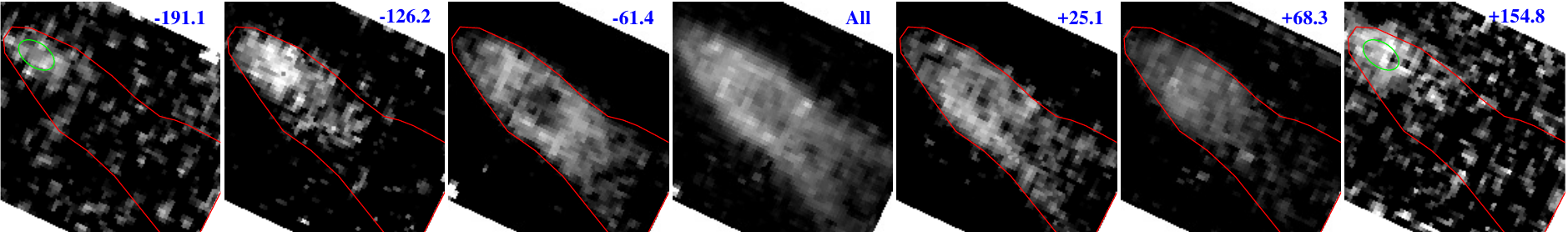} \\
\caption{GMOS-N IFU H$\alpha$ velocity channel images, covering $5.5^{\prime\prime} \times 6.1^{\prime\prime}$. The combined IFU image from 2016 is shown in the middle panel. This is flanked by velocity slices (central velocity, in km/s, at upper right in each frame), with the H$\alpha$ limb as a red outline, for comparison. The oval region near the apex provides the line spectrum shown in Figure \ref{fig:IFUspec}.}
\label{fig:IFUchan}
\end{figure*}
\section{Observations and Data reduction}
\label{sec:data}

\subsection{Chandra Observations}
\begin{table}
\begin{flushleft}

\caption{Overview of B2224 \textit{CXO} observations used in this paper. The `Aim' column indicates aimpoint on the ACIS-I or ACIS-S array. `Exp' lists the exposure times in kiloseconds, after filtering out periods of high background.}
\setlength{\tabcolsep}{3pt}

\begin{tabular}{cccc|cccc}
\hline \hline
Date & Obs & Aim & Exp & Date & Obs & Aim & Exp \\
 & & & [ks] &  & & & [ks] \\ \hline
2000-10-21 & 755 & S& 48.8 &  2021-04-21 & 24433 & I & 25.7 \\
2006-08-28 & 6691& S& 10.0 & 2021-04-23 & 24431 & I & 25.7\\
2006-10-06 & 7400& S& 36.6 & 2021-04-25 & 24429 & I & 24.5\\
2012-07-28 & 14467& S& 14.6& 2021-07-04 & 24428 & I & 29.7\\
2012-07-29 & 14353& S& 34.6 & 2021-07-25 & 24427 & I & 24.7\\
2012-08-01 & 13771& S& 49.2 & 2021-10-09 & 24430 & I & 29.7\\
2021-02-19 & 24437 &I& 24.7 &  2021-10-20 & 23537 & I & 57.2\\
2021-03-14 & 24434 & I &29.5  & 2021-11-14 & 24432 & I & 29.6\\
2021-03-15 & 24435 &I& 14.6 & 2022-02-21 & 24426 & I & 20.9\\
2021-03-16 & 24992 &I & 14.9 & 2022-02-24 & 26336 & I & 18.0 \\
2021-04-03 & 24436 & I & 24.3 \\ \hline

\end{tabular} \\
\label{tab:ChandraObs}
\end{flushleft}
\end{table}

The Guitar, at its high Northern declination, requires short dwell times for the thermal health of {\it CXO}. Thus we obtained fifteen $15-60\,$ks visits between 2021 February 19 and 2022 February 24 to collect 393.8 \,ks of exposure. For all observations we used the I3 chip of ACIS-I array, with the aimpoint positioned so that the full filament was covered at any roll angle. An overview of the new and archival observations is given in Table \ref{tab:ChandraObs}. In addition to the 2021-2022 epoch of 393.8 ks, we re-analyze the three archival epochs: 2000.89 (48.8ks total, Ep1), 2006.80 (46.5 ks total, Ep2), and 2012.66 (98.4 ks total, Ep3). 

All data were reprocessed with the standard CIAO reprocessing tools, using CIAO 4.12 and CALDB 4.9.1. We performed a relative astrometric correction by using \textit{wavdetect} on the aimpoint chip of each observation (S3 for the archival epochs, I3 for the latest epoch) and then using \textit{reproject\_aspect} to minimize the point source offsets between observations. ObsID 23537 was used as the reference observation, because it has the longest exposure time in the new epoch. The pulsar was excluded from the point source list, because its large proper motion makes it unsuitable for astrometry. For most ObsIDs, at least 4 sources could be used for registration. However the degraded soft X-ray response of ACIS and the short exposure time in the later observations (most notably 24435, 24436, 24992, and 26336) combined with the lack of large numbers of bright field stars means that for these obervations only 2-3 matching sources could be found. The bright source directly NW of the PSR unfortunately decreased in brightness over the course of 2021, making this source difficult to use for astrometry in the later observations of the 2021 epoch as well. We estimated the error on the frame registration from the RMS residuals after source matching, and summed the errors of individual frames weighted by exposure time to find the error on the relative astrometric correction in each epoch. We find $1 \sigma$ errors of 0.08, 0.12, 0.11, and 0.16 arcsec for the 2000, 2006, 2012, and 2021-2022 epochs respectively. After the astrometric correction, we combined the event files for each epoch and generated $1-5$\,keV exposure maps using a power law with $\Gamma=1.5$ (appropriate to the PWN emission, see \S3.3) as an input energy weighting, with the tool \textit{merge\_obs}.

In our X-ray analysis, we compare the exposure-corrected data from the S and I chips, which have significantly different particle background levels. We therefore estimated the particle background  contribution by retrieving the ACIS `stowed' background maps from the calibration database, and scaling them by the number of 9.5-12\,keV counts of the observation in question (no bright sources were present to contribute significant photon counts to this band). The scaled backgrounds were subtracted from the data before making the exposure-corrected image for each epoch.

\subsection{New {\it HST} Apex Image}
\label{sec:HST}

To probe the current state of the bow shock we obtained a new {\it HST} ACS H$\alpha$ image of the nebula apex (the `head' of the Guitar) under Program 16426. This is a challenging observation, since, although the head is the brightest portion of the nebula, its surface brightness is still low. And while the ACS/WFC has the highest H$\alpha$ sensitivity of the present {\it HST} cameras, at low light levels it suffers severely from a degraded Charge Transfer Efficiency (CTE). Happily at $DEC=+65^\circ$, the Guitar lies far enough North to be occasionally in the Continuous Viewing Zone and CVZ observation were kindly granted for the three awarded orbits by {\it HST}, allowing much longer exposures, and greater photoelectron count per pixel at readout, than would otherwise be available. In the end we were able to schedule five 2910\,s F658N exposures (and two 338\,s F625W frames for continuum monitor and subtraction) on 2020 November 8 (MJD 59161). With few exposures we also reduce the total read-noise cost of the observation but slightly decrease the CR rejection efficiency. The second mitigation is to place the Guitar apex near the WFC1-CTE position, where the number of row transfers is minimal, decreasing CTE degradation. The cost is that WFC field of view is cropped closely around the apex, decreasing the number of upstream field star detections for registration and context. In addition, the geometrical distortions are largest near the array corners, such as the CTE1 position. Happily, at the observation epoch the default roll angle placed the body of the Guitar farther on to the WFC array. Its surface brightness is very low so that the only useful measurements of the body at {\it HST} resolution are the limb position in some of the brighter areas. Nevertheless it is gratifying to detect the whole structure (Figure \ref{fig:largescale}).

In the end these mitigation measures were successful and we have obtained the best-ever image of the Guitar head. Since the pulsar had moved $2.74^{\prime\prime}$ since the last {\it HST} exposure (\S3.1), there are quite substantial changes. All the {\it HST} data used in this paper can be found in MAST:\dataset[10.17909/ytdx-cf49]{http://dx.doi.org/10.17909/ytdx-cf49}.

\subsection{Archival IFU apex data}

Inspecting the Gemini archive, we found unpublished GMOS-N IFU observations of the Guitar nebula taken on 2008 June 10 and July 3 (Program GN-2008A-Q-3, van Kerkwijk, PI). The observations included $4\times 3863$\,s exposures with the IFU-2 mask, the B1200 grating and RG610 filter, covering the head of the nebula with a $0.2^{\prime\prime}$ fiber grid. A few 300\,s direct acquisition images using the G0310 H$\alpha$ filter and associated calibration files were also obtained. Conditions were good, meeting the 20\% best seeing criterion. 

To improve S/N on this faint nebula, the GMOS-N detector was binned $2\times2$ during this observation. As it happens, the $2\times$ spatial binning left the fiber traces poorly resolved. This is generally not recommended and, indeed, meant that the Gemini pipeline software failed to trace the spectra and extract the data. To complete reductions, we therefore had to mark the 1000 fiber spectra positions by hand near the position of the H$\alpha$ feature in each spectral exposure flat field image and force a low-order trace centered on these fiber centroids. Importing these traces to the arcs we were able to use line features (again initially identified by hand) to establish a good wavelength solution. The traces applied to the target integrations could then be used to extract and calibrate the 1-D spectra. These were spatially flattened using sky lines and assembled into position-velocity cubes. The final weighted combination of these data gave a data cube with a spatial scale of $0.1^{\prime\prime}$/pixel and a wavelength scale of 0.4729\AA/image plane (21.6 km/s/image plane). The spectral resolution of 1.23\AA (56\,km/s) was confirmed by measuring sky lines and the $0.6^{\prime\prime}$ FWHM spatial resolution of the pre-images is maintained in the data cubes, as indicted by the width of the narrow H$\alpha$ limb $\sim 2.5^{\prime\prime}$ behind the apex. 

\begin{figure}
\centering
\includegraphics[width=3.2in]{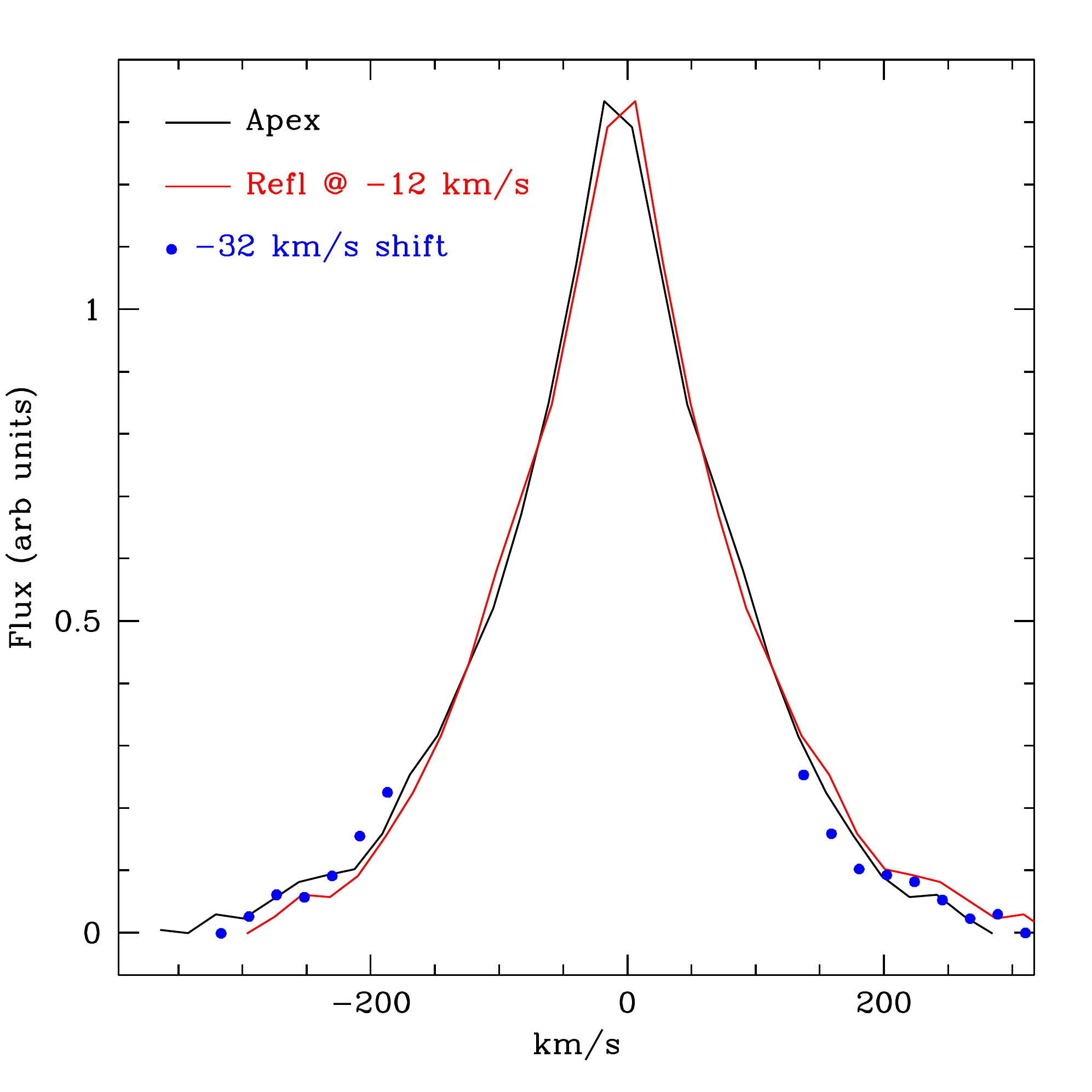} \\
\caption{Velocity profile just behind the nebula apex (oval in Figure \ref{fig:IFUchan}). A reflected version of the profile (red) shows that it is centered at $\sim -12$\,km/s. The high velocity $|v|=150-300$\,km/s wings (blue points, shifted from the red profile to match original apex profile wings) show a slight additional blue shift. }
\label{fig:IFUspec}
\end{figure}

We find that the nebula is best detected in low velocity channels near the nebula limb, as expected from neutral excitation in the post shock gas (and by projection effects). Larger velocities are found principally right behind the nebula apex, where charge exchange allows accelerated post-shock protons to obtain electrons and emit H$\alpha$. The line spectrum from a region just behind the  apex is shown in Figure \ref{fig:IFUspec}. The similarity of the channel maps red and blue of the central velocity supports the inference from prior bow shock image fits that the pulsar space velocity lies close to the plane of the sky. Focusing on the velocity extrema we see an offset of $-32-(-12)=-20$\,/km/s for the wing components peaking at $\sim \pm 250$\,km/s, giving $\theta_v \approx {\rm arctan}(-20{\rm km/s/250 km/s})\approx 5^\circ$ out of the plane of the sky. 

Assuming approximate axial symmetry, we can check these velocities by examining the transverse expansion of the head boundary in the HST images. By comparing the 2020 and 2006 HST images we can see that across the bulk of the head region, starting $5^{\prime\prime}$ back from the 2020 apex, the lateral expansion is $\approx 0.3^{\prime\prime}$ over 14y or $\approx \pm 85 {\rm km/s}$ at $d=0.83$\,kpc, in good agreement with the brightest emission in the channel maps. It is a bit more difficult to measure the transverse expansion corresponding to the elliptical apex region marked in Figure \ref{fig:IFUchan}, but comparing the 2008 GMOS-N pre-image with the 2006 HST frame, we see transverse expansion of $\sim 0.1^{\prime\prime}$ over 1.7y or $\sim \pm 230 {\rm km/s}$, in reasonable agreement with the wing component speeds in Figure \ref{fig:IFUspec}.

\section{Multi-Epoch Comparison}

Over the past 25 years we now have 4 epochs of observation each with {\it HST} in H$\alpha$ and {\it CXO} in X-rays (Table \ref{tab:ChandraObs}). Other ground-based H$\alpha$ images exist, but the very small angular scale of the bow shock stand-off requires {\it HST} resolution for serious study of the evolving morphology. Comparison of the shock between these epochs gives important clues to the nature of the filament.

\begin{figure*}
\hspace*{-5mm}\includegraphics[scale=0.5]{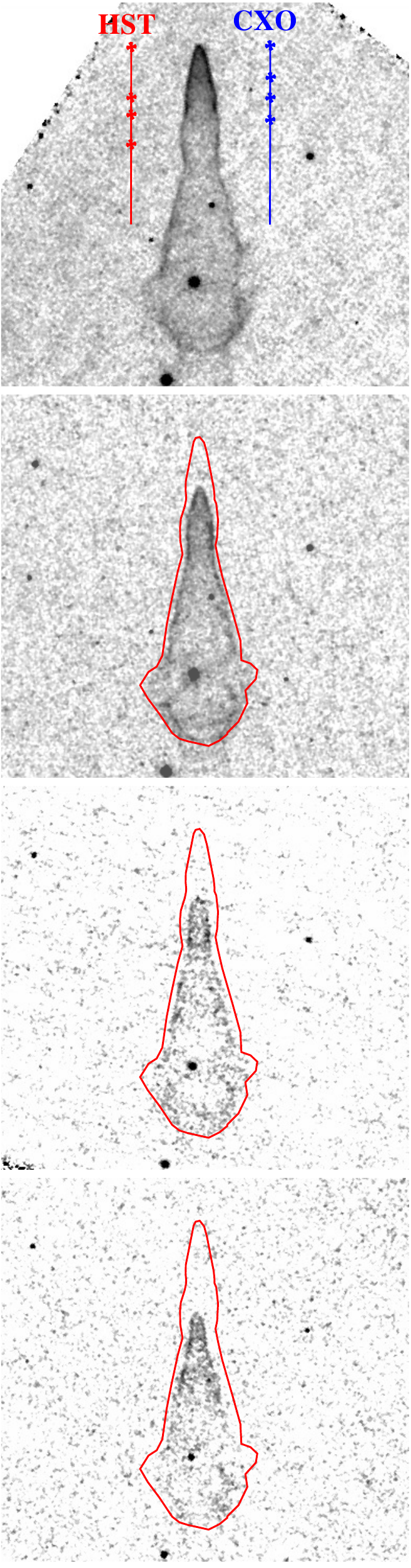}
\vskip -13.2cm\hskip 27mm \includegraphics[scale=0.82]{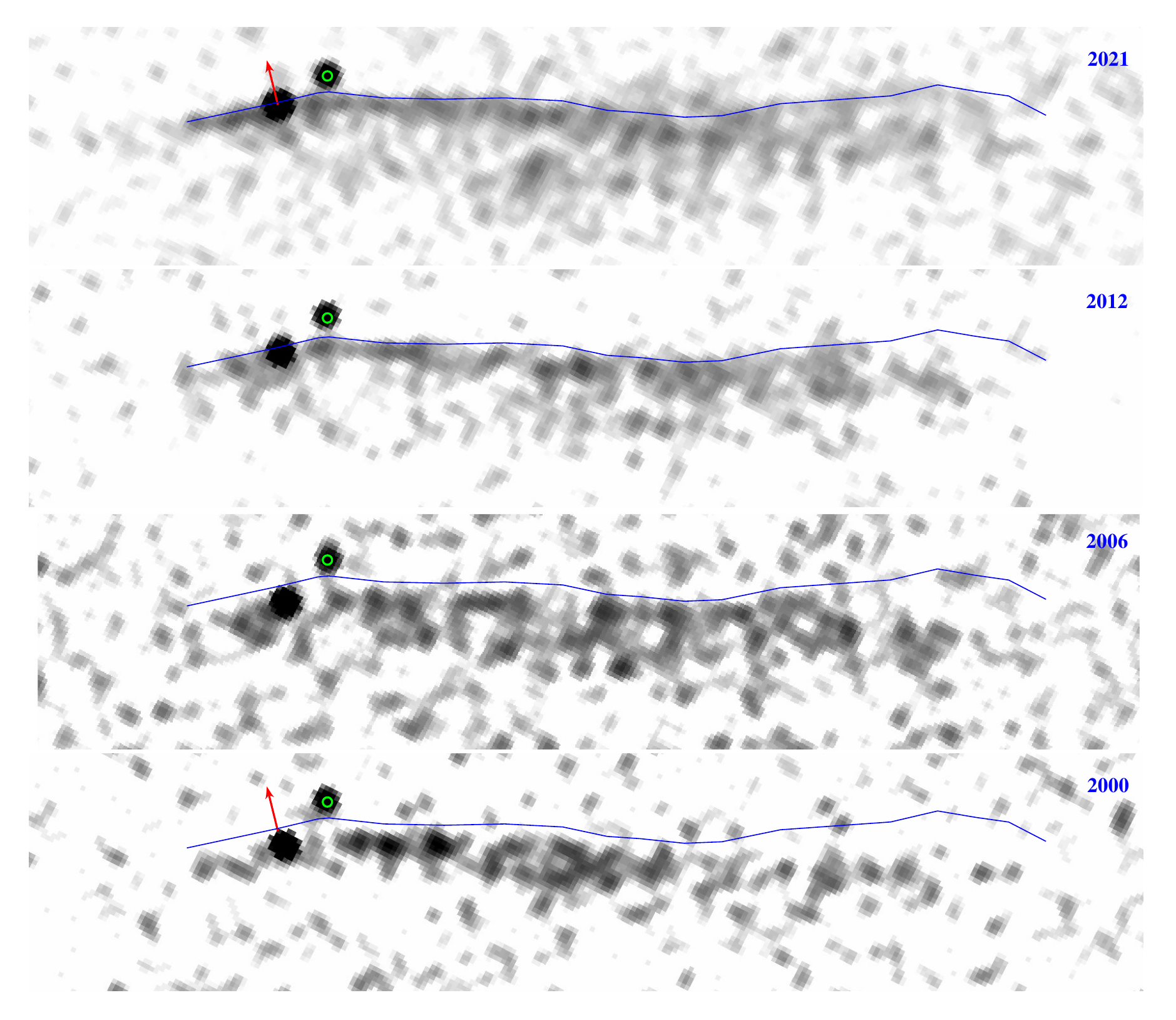}
\vskip -0.4cm
\caption{1994-2021 Guitar/Filament Evolution. Left panels show four $22^{\prime\prime}\times 21^{\prime\prime}$ cutouts from the {\it HST} H$\alpha$ images (1994, 2001, 2006, 2020). The top panel has the pulsar positions at the four {\it HST} and four {\it CXO} epochs marked on lines representing 50y ($9.7^{\prime\prime}$) of proper motion. The lower panels have the outline of the 2020 {\it HST} H$\alpha$ limb marked, for comparison. The right panels show the filament evolution over the four {\it CXO} epochs ($4^\prime \times 1^\prime$ cutouts). The filament leading edge from the 2021 epoch is marked by the blue line, while the red arrow in the top and bottom panels shows the 50y proper motion, and a stationary background source is marked for reference.
}
\label{fig:4ep}
\end{figure*}

\subsection{Optical Evolution}

{\renewcommand{\arraystretch}{1.15}

\begin{table}
\centering
\caption{The standoff distances (in mas) measured for each \textit{HST} epoch, for thin and wide shock models; shown are the median values from the posterior distributions, with the 14th and 86th percentiles as the errors}. The estimates of \cite{Chatterjee2004} and \cite{Ocker2021} are also shown. 
\begin{tabular}{c c c c c }
\hline \hline
Epoch & $r_{0,\rm{thin}}$& $r_{0, \rm{wide}}$& $r_{0,\rm{O2021}}$& $r_{0, \rm{CC2004}} $\\ \hline
1994 & $86_{-9}^{+12}$ & $94_{-7}^{+11}$ & $77 \pm 4$ & $120 \pm 40$ \\
2001 & $116_{-15}^{+16}$ & $112_{-14}^{+14}$ & $110 \pm 10$ & $100 \pm 40$ \\
2006 & $97_{-4}^{+4}$ & $93_{-3}^{+3}$ & $94 \pm 6$ & \\
2020 & $96_{-2}^{+4}$ & $92_{-2}^{+3}$ &  \\ \hline
\end{tabular}
\label{tab:standoff}
\end{table}}

Comparing the optical images at the left hand side of Figure \ref{fig:4ep} with the line marking the limb of the 2020 image we see that as the pulsar advances, the perpendicular expansion is rapid at the apex but slows by a few arcsec behind. This is also visible in the IFU data cube. The general structure of the Guitar head is best seen in our new high S/N 2020 image; it is roughly symmetric, with indentations, especially a `pinch' $\sim 7^{\prime\prime}$ behind the pulsar, and higher limb brightness regions, e.g.\,$\sim 3^{\prime\prime}$ and $\sim 6^{\prime\prime}$ behind the pulsar. Thus the geometry of the bow shock apex and its expansion rate must vary. The most extreme illustrations of this are, of course, the apparently closed bubble of the Guitar head and the double cavity of the Guitar body itself. The approximate bi-lateral symmetry of the overall nebula indicates either that the central pulsar wind varies or that the perturbations producing these structures have a coherence scale substantially larger than the width of the nebula. However, there is also significant right-left asymmetry, which implicates instabilities in the shock flow or variations in the external medium on the few arcsec scale of the head width. 

\begin{figure*}
\includegraphics[width=.95\textwidth]{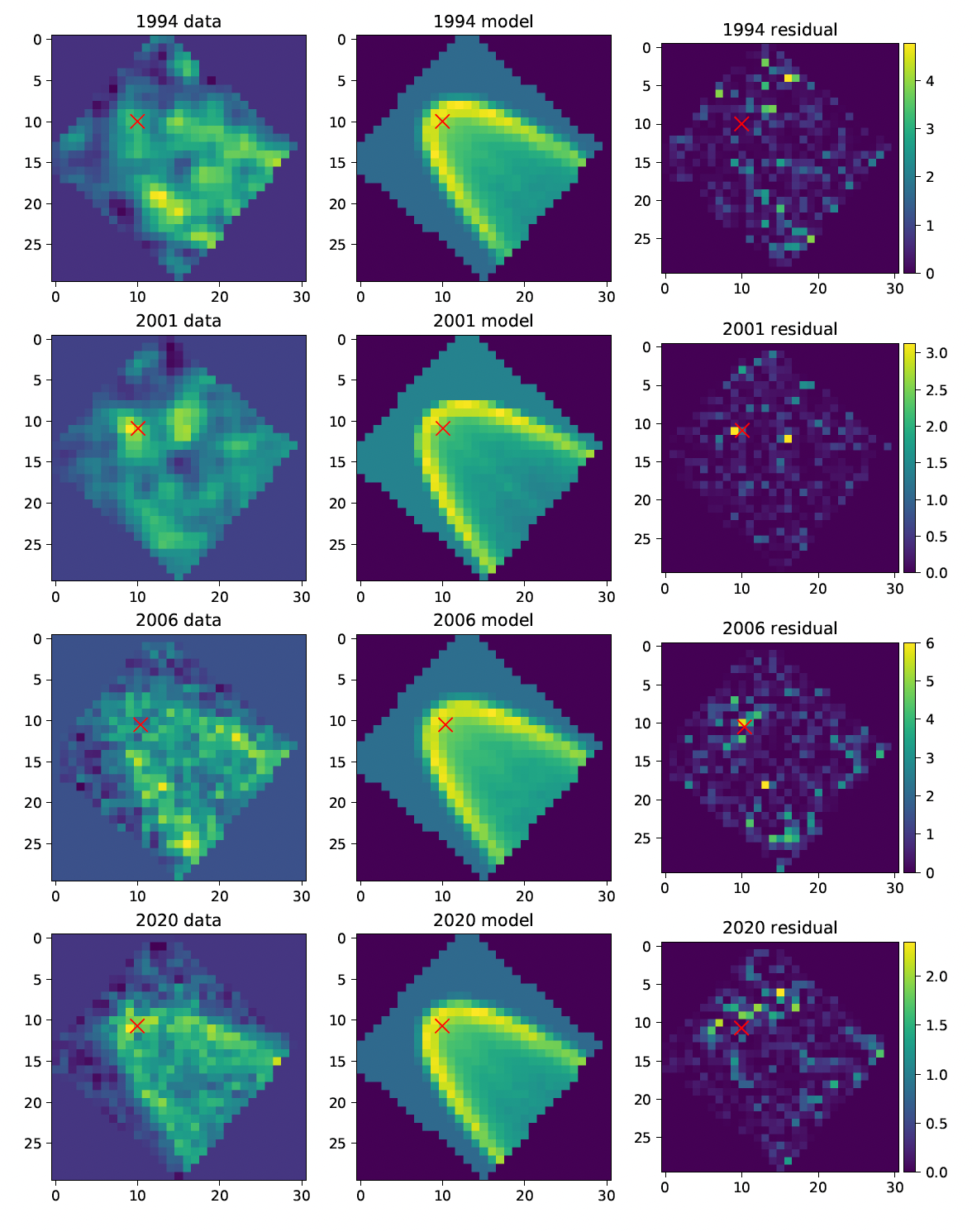} 
\caption{The Wilkin-model bow shock fits for the 4 different H$\alpha$ epochs. All images are shown in native resolution of $0.05\arcsec$/pixel. The left column shows a cut-out of the bow shock apex for the 4 epochs, using pixels up to $\approx 1^{\prime \prime}$ behind the apex. The 1994 and 2001 epochs have been lightly smoothed for visualization. The middle column shows the Wilkin model, and the rightmost column shows the residual (data-model)$^2$/$\sigma^2$. The red cross shows the location of the pulsar - a free parameter for the 2020 epoch, and referenced to Gaia using a set of reference stars for the other three epochs.}
\label{fig:apexfit}
\end{figure*}

\begin{figure*}
\includegraphics[width=.95\textwidth]{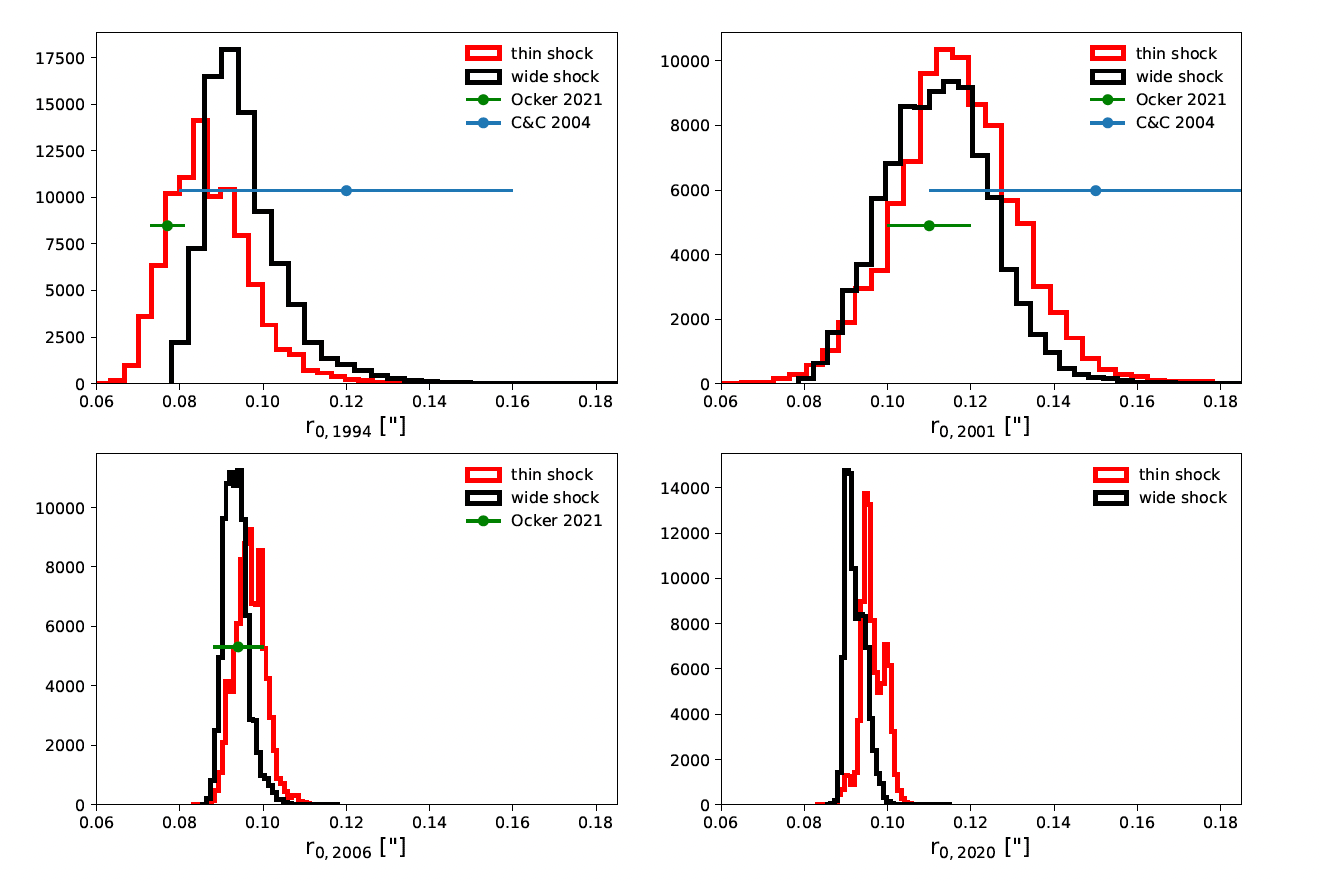} 
\caption{The posterior distributions for the standoff distance $r_0$ in each epoch (left to right, top to bottom: 1994, 2001, 2006, 2020. Shown for reference are the estimated standoff distances of \citet{Ocker2021} and \citet{Chatterjee2004}}
\label{fig:posteriors}
\end{figure*}

The spectrum of such perturbations have recently been explored by \citet{Ocker2021}, who, following \citet{Chatterjee2004} discuss apparent changes in the bow shock standoff distance in the previous three {\it HST} epochs. These $\theta_0$ were, however, estimated by marking the apparent bow shock limb by hand and then fitting to these marked points. This, of necessity, introduces substantial subjectivity. We therefore have sought to fit \citet{Wilkin2000} apex models directly to the {\it HST} images. This model computes the locus of the contact discontinuity, which for a `thin' shock marks the H$\alpha$ front. In practice post-shock pressure widens the structure; the H$\alpha$ emission standoff should be $\sim 1.3 r_0$ at the apex and this factor should grow downstream. A simple approximation increases the transverse scale by $1.25\times$ \citep{2014ApJ...784..154B} for a `wide' shock model. For the first three epochs we were able to register the frames to \textit{Gaia} stars to determine the position of the pulsar in the frame with an $1 \sigma$ uncertainty of 0.07, 0.07, and 0.06 pixels respectively. For 2020, however, WFC corner distortions defeated such registration, so we have let the pulsar position adjust over a 1 pixel ($\sim 50$mas) range. We used the affine invariant Markov Chain Monte Carlo (MCMC) algorithm of \citet{Goodman2010}, implemented through the python package {\sc emcee}, to sample the likelihood function and obtain posterior distributions for the standoff distance in each epoch. The MCMC routine was run using 50 walkers and 5000 steps for each walker. In order to run MCMC efficiently, we first performed a maximum likelihood analysis, and started the walkers with initial parameters close to the best-fit parameters. Upon visual inspection of the chains, we further excluded the first 500 steps of each walker to `burn in' the chains, ensuring convergence. We calculate the integrated autocorrelation time $\tau_f$ to be $
\sim 50$ steps, meaning that the walker length of 4500 steps should be sufficient.\footnote{In the {\sc emcee} documentation on autocorrelation analysis (\url{https://emcee.readthedocs.io/en/stable/tutorials/autocorr/)}, it is suggested that each walker should have a length of $>\,50\tau_f$ steps, so that enough independent samples can be obtained to yield accurate results.}.A visual comparison of the data and the Wilkin thin shock model for each epoch is shown in Figure \ref{fig:apexfit}. The posterior distributions for $r_0$ are plotted in Figure \ref{fig:posteriors}, with values listed in Table \ref{tab:standoff}. For the 2020 epoch, the position uncertainty from the posterior (the 14th and 86th percentiles of the distribution) is $\approx 0.15$ pixel ($\sim 8 $\,mas) in both the $x$ and $y$ directions.

The \citet{Ocker2021} estimates generally lie within the $r_0$ uncertainty ranges, but have nominal errors much smaller than we find for a direct fit, especially for the 1994 and 2001 data. Alas our more realistic errors mean that direct evidence for $r_0$ variation is poor. The bulk of our uncertainty range suggests that $r_0$ was larger in 2001, but even this result is weak. Additional images of the quality of our new ACS/WFC exposure are needed to probe stand-off variation at the required $\sim 5$mas level. 

Nevertheless the head limb shape does suggest that the standoff was small when pulsar was at the position of the head's closed base, $\sim 15.5^{\prime\prime}$ behind the present apex (i.e. in $\sim 1940$). The transition into the head bubble may be similar to the `break-through' inferred for the PSR J2030+4415 H$\alpha$ nebula and filament \citep{deVries2022}. The `pinch' $\sim 7^{\prime\prime}$ back (i.e. in 1985), and the increased limb brightness $\sim 3.2^{\prime\prime}$ behind the apex (in $\sim 2004$) suggest weaker compression events.

\begin{figure}
\includegraphics[width=0.48\textwidth]{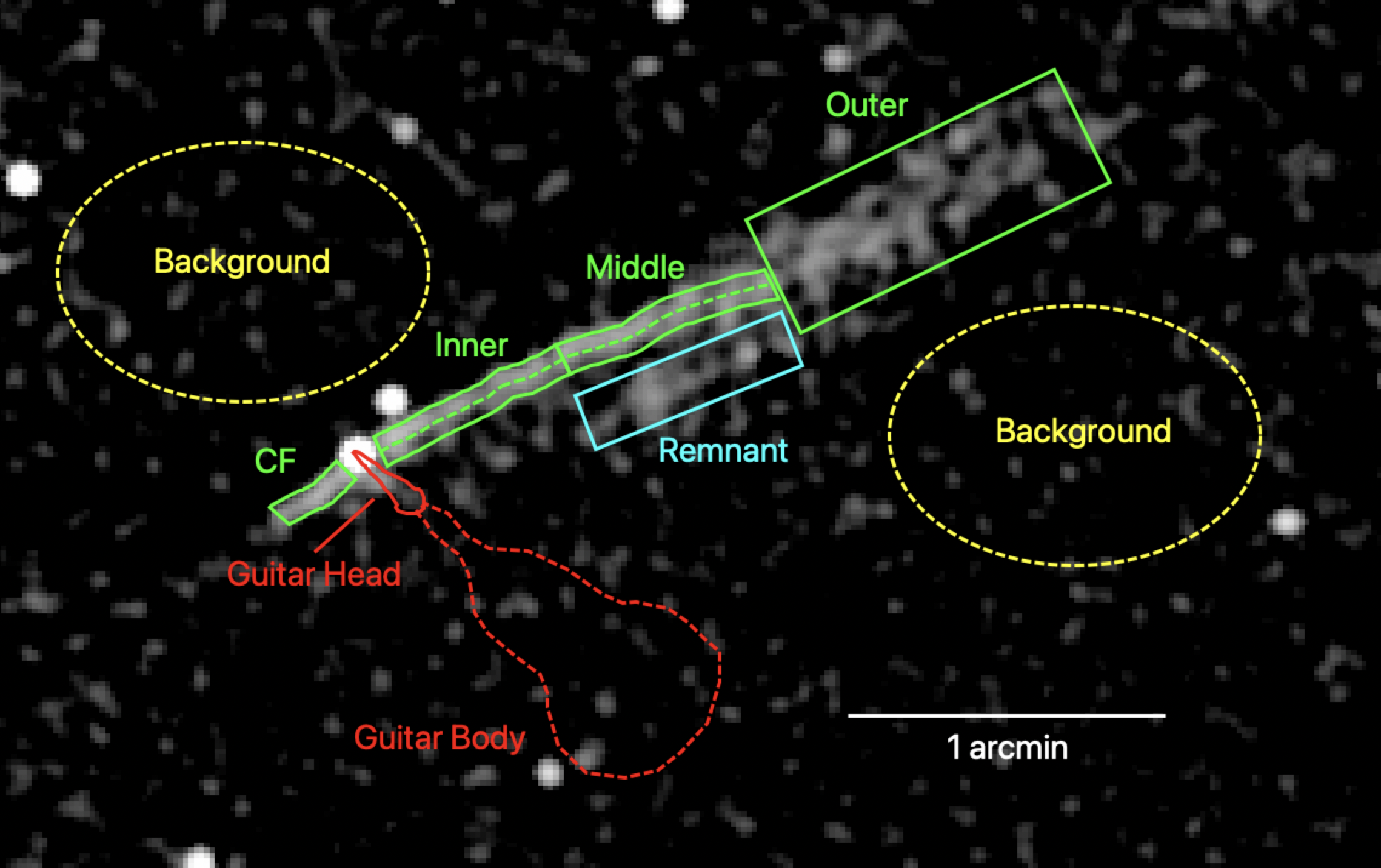} 
\caption{Regions used for spectral analysis of the filament in the 2021 \textit{Chandra} epoch. We divided the filament into three main sections: the inner section where the leading edge is sharpest (0--0.7$^\prime$), the middle section where the filament appears to become more diffuse (0.7--1.3$^\prime$) and the outer, most diffuse section where the sharp leading edge has largely disappeared (1.3--2.4$^\prime$). Additionally, the inner and middle sections are divided by the green dashed line into the 'Leading' and 'Trailing' regions. We also identify a `Remnant' region of bright emission around $12\arcsec$ behind the leading edge in the middle section. The red regions show the contours of the Guitar nebula head (solid line) and body (dashed line) in the 2020 \textit{HST} H$\alpha$ image. }
\label{fig:specregs}
\end{figure}

\subsection{X-ray Evolution}


\begin{figure}
\centering
\vskip -3mm
\includegraphics[width=0.49\textwidth]{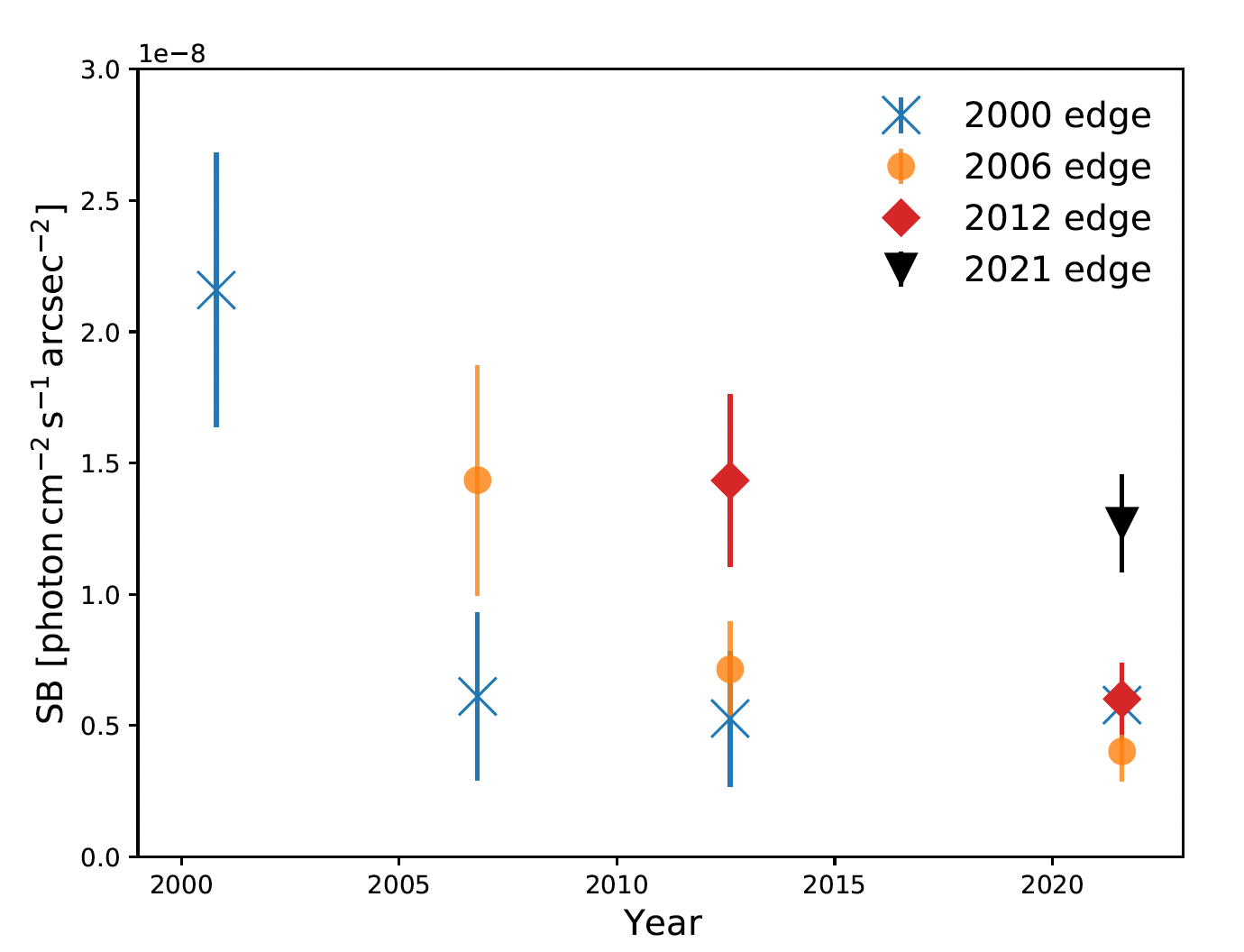} 
\vskip -3mm
\caption{Light-curves of the leading edge ($1^{\prime\prime}$ width) of the inner filament (0--0.7$^\prime$ segment), showing the flux in each epoch and the flux of that same region of the sky in following epochs. The y-axis indicates the 1-5 keV photon surface brightness.}
\label{fig:fil_lc}
\end{figure}

\begin{figure}
\centering
\vskip -3mm
\includegraphics[width=0.49\textwidth]{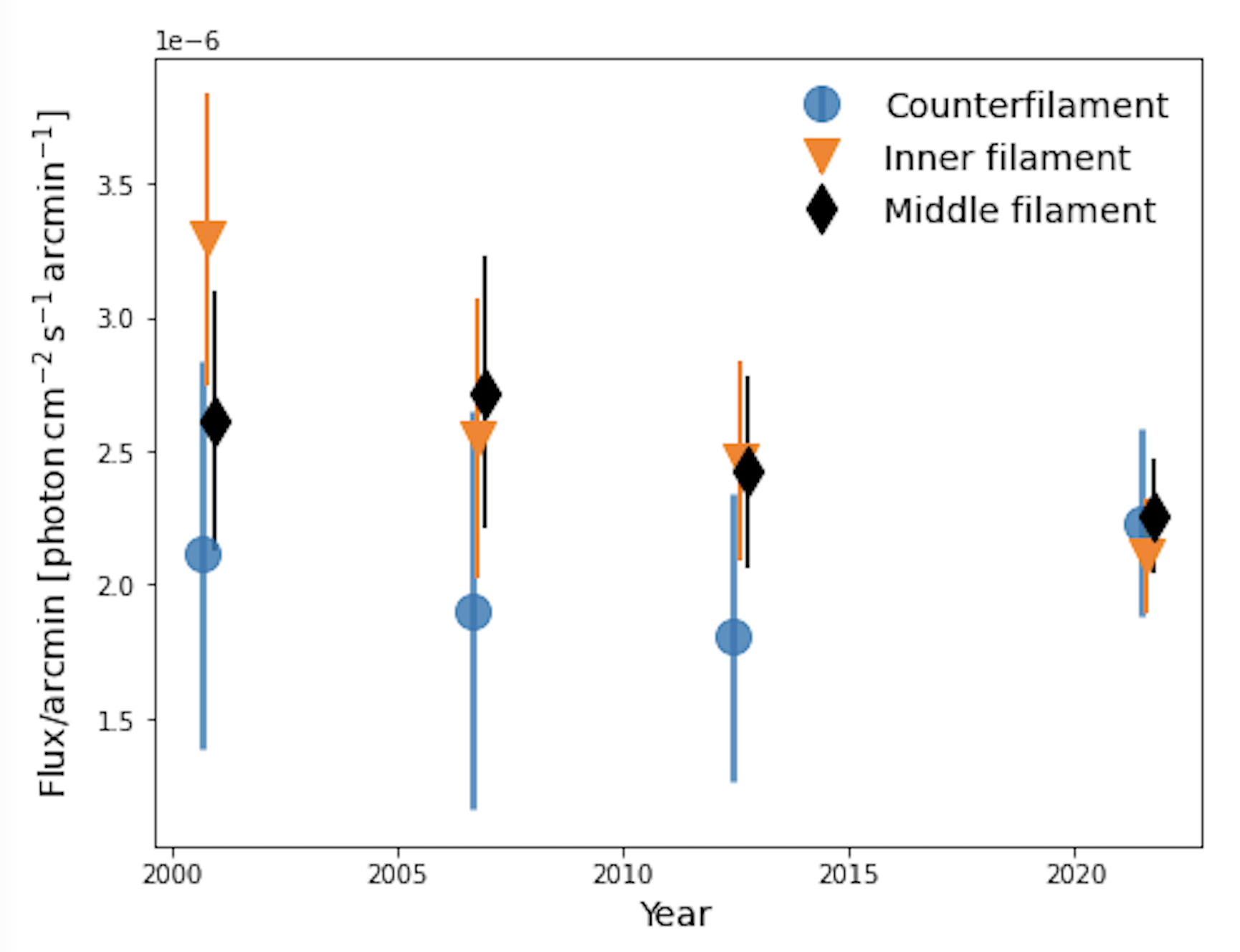} 
\vskip -3mm

\caption{Light-curves of the inner counter-filament and inner and middle filament sections. Data points have been slightly offset from each other for legibility. The y-axis indicates the 1-5 keV photon flux per arcmin length of filament. The lengths of the inner counter-filament, and inner and middle filament sections are 0.25$^\prime$, 0.65$^\prime$, and 0.68$^\prime$ respectively and fluxes have been integrated across the width of the main filament.}
\label{fig:fil_lc_wide}
\end{figure}

In Figure \ref{fig:specregs} we define several regions useful in describing the filament's spectrum and its evolution.
The $e^\pm$ injection site shifts with the steady pulsar motion, and in Figure \ref{fig:4ep} it is apparent that the filament leading edge marches along with the pulsar, as also noted by \citet{2021RNAAS...5....5W}. Our deeper 2021 exposure provides a much better view of the counter-filament (CF) than earlier epochs. It extends at least $20^{\prime\prime}$ and likely $40^{\prime\prime}$ from the pulsar. Interestingly it does not line up well with the filament leading edge, instead intersecting the proper motion axis some $1.5^{\prime\prime}$ behind the pulsar position. Both it and the filament have substantial curvature near the bow shock. This is likely a field line `draping' effect or field distortion from supra-thermal particles as most clearly seen in the `lighthouse' PWN filament and counter-filament \citep{2016A&A...591A..91P}.

We checked to see if the PWN PSR trail is detected in our deep 2021 image. Using the `head region' of the Guitar (see Figure \ref{fig:specregs}) as an aperture and subtracting similar flanking regions as background, we find an excess of $11\pm 4.5$ counts in the 0.7--5\,keV, range, a marginal $2.5\sigma$ detection. This gives a filament/trail flux ratio $>100$, the largest among known filaments.

In Figure \ref{fig:fil_lc} we measure the surface brightness at the filament leading edge in each epoch and compare the flux in the same aperture in subsequent epochs. In general the region corresponding to the edge shows an initial rapid decrease in surface brightness in the following epoch (see also Figure \ref{fig:4ep}). We infer a rapid change in the electron population as the pulsar moves ahead to the next set of field lines, due to cooling, advection or diffusion. The subsequent brightness decrease, if any, is much smaller. Fortuitously the leading edge was much brighter than usual during the original 2000 epoch, which helped in the filament's discovery. This may be related to enhanced injection around this epoch. For example if $r_0$ decreases, then more pulsar/PWN shock particles have gyroradii exceeding $r_0$, so escape to the filament might increase and the filament surface brightness may temporarily increase. Averaged over the full width the fluxes per unit length seem quite constant (Figure \ref{fig:fil_lc_wide}); although the inner counter-filament appears more prominently in the 2021 image, its flux per unit length remains consistent with that of the filament, within errors.

\begin{figure}
\centering
\vskip -1.9mm
\includegraphics[width=0.49\textwidth]{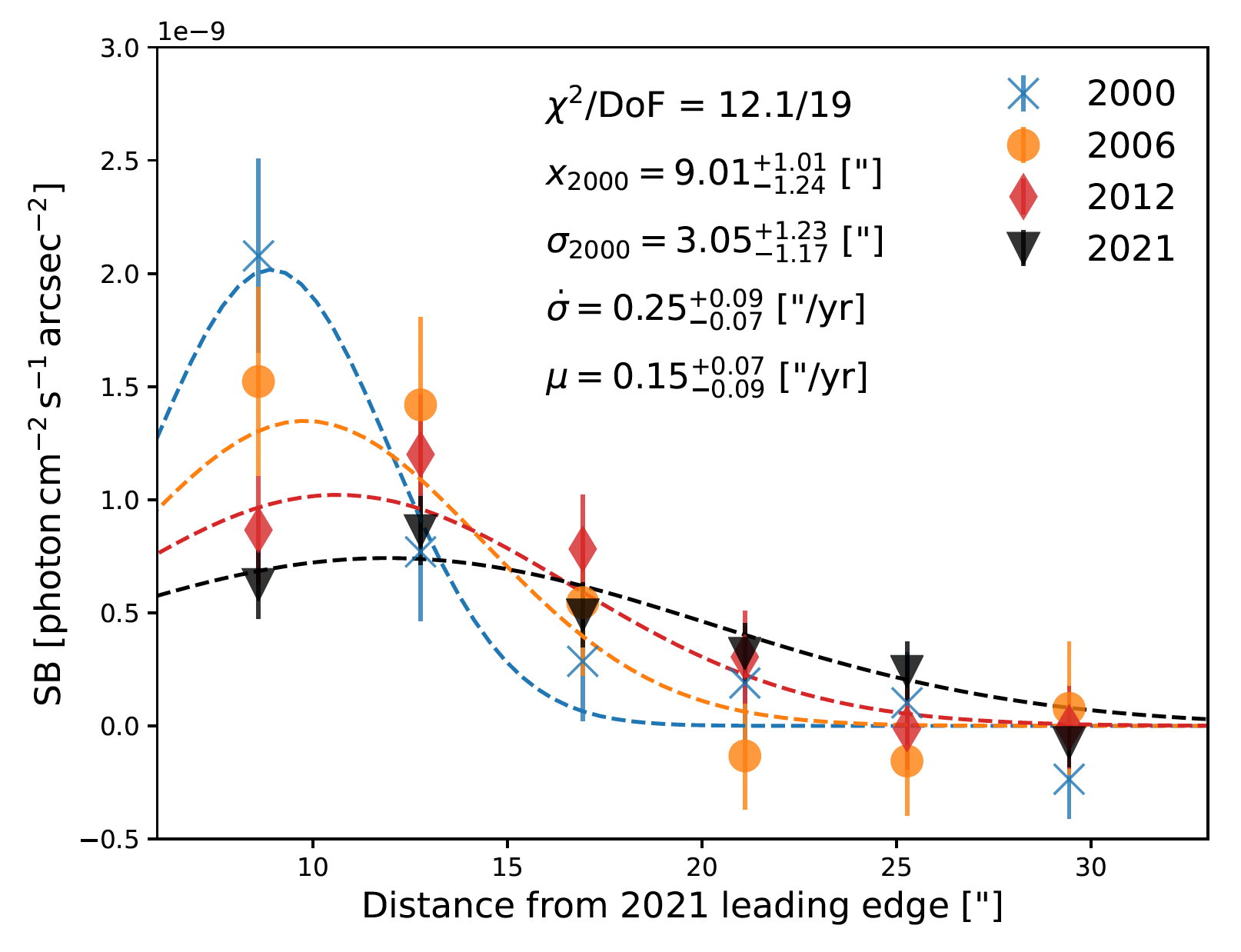} 
\vskip -3mm
\caption{Spread of the emission behind the leading edge in the filament middle section across four epochs. The data are well described by a steady shift of the peak behind the leading edge, a steady increase in the width, and a constant integrated flux. The y-axis indicates the 1--5 keV photon surface brightness. Labels indicate the best-fit parameter value (with $1 \sigma$ errors ): $x_{2000}$ and $\sigma_{2000}$ are the peak position and standard deviation of the Gaussian component in 2000 respectively; $\dot{\sigma}$ indicates the increase in $\sigma$ over time; and $\mu$ indicates the shift of the peak away from the leading edge over time.}
\label{fig:spread}
\end{figure}

Although the statistics are limited in the early images, there appear to be changes behind the leading edge. In Figure \ref{fig:4ep} the most notable changes are in the `Middle' section of the filament where the emission spreads behind the leading edge as a shifting ridge. We quantify this trend in Figure \ref{fig:spread}, where fits to Gaussian distributions transverse to the filament show a progressive shift and broadening of the maximum. Note that the integral flux is consistent with constant across the four epochs. A fit to such regions in the `Inner' zone gives similar evolution with nearly identical parameters, but lower statistical significance. We attempt to interpret these results in the conclusions.

\subsection{Spectral fits}

We have extracted spectra for the several regions of Figure \ref{fig:specregs} using the standard \textsc{CIAO} tools. To each of the spectra, we have fit a power law multiplied by Galactic absorption, which we have set at $2.7 \times 10^{21}$ cm$^{-2}$. The results of the spectral analysis are shown in Table \ref{tab:specfits}. There are no significant differences in spectral index between the regions. The weak evidence for spectral softening with distance from the pulsar would require much deeper observation for a serious test.

Additionally, we have estimated the magnetic field strength under the assumption of equipartition. For an optically thin region filled with relativistic electrons and magnetic field emitting synchrotron radiation
\begin{equation}
\label{eq:syncB}
B = 46 \left[ \frac{J_{\rm -20}(E_1,E_2) \sigma } {\phi}  \frac{C_{1.5-\Gamma}(E_m, E_M)}{C_{2-\Gamma}(E_1,E_2)}\right] ^{2/7} \mu G
\end{equation}
where 
\begin{equation}
C_q(x_1,x_2) = \frac{x_2^{q} - x_1^q}{q}.
\end{equation}
$J_{\rm -20}(E_1,E_2) = 4 \pi f_{\rm -20}(E_1, E_2) d^{2}/ V$ is the observed emissivity (in $10^{-20}$\,erg\,s$^{-1}$\,cm$^{-3}$, between $E_1$\,keV and $E_2$\,keV), $\sigma=w_B/w_e$ is the magnetization parameter, $\phi$ the filling factor, and $E_m$ and $E_M$ the minimum and maximum energies, in keV, of the synchrotron spectrum with photon $\Gamma$. We assume that the structures are cylindrical, with diameter set to the observed region width. We list the derived equipartition fields in Table \ref{tab:specfits} for $\sigma=\phi=1$, $E_m=0.01\,\rm{keV}$ and $E_M=10\,\rm{keV}$.

\begin{table}
\setlength{\tabcolsep}{3.5pt}
\caption{Spectral fit results (with $1 \sigma$ errors) for the filament in the 2021 \textit{Chandra} epoch (see Figure \ref{fig:specregs} for the regions). The `Leading' and `Trailing' regions are composed of the front and back halves, respectively of the combined inner and middle regions.  B$_{eq}$ was computed assuming cylindrical volumes for each region.}
\centering 
\label{tab:specfits}
\begin{tabular}{l l l l l l} %
\hline\hline 
Region		& Counts &$\Gamma$ &$f_{-15}$\textsuperscript{b}& $\chi^2$/DoF & B$_{eq}$  \\
& & & & & [$\mu G$] \\ [0.5ex]  \hline 
Inner & $214\pm17$ & $1.31\pm0.16$ & 9.9 &29.3/27 & 13 \\
Middle & $209\pm17$ & $1.37\pm0.17$ & 10.2 & 24.3/24 & 14 \\
Outer & $489\pm32$& $1.58\pm0.15$ & 24.1 &  53.2/48 & 8 \\
CF & $86\pm11$& $1.71\pm0.30$ & 3.5 & 23.8/24 & 17 \\
Leading & $273\pm19$ &  $1.39\pm0.14$ &  13.6 & 22.7/33 & 19 \\
Trailing & $154\pm16$ & $1.60\pm0.20$ & 7.1 & 30.5/30 & 17 \\
Remnant &  $174\pm19$ & $1.40 \pm 0.27$  & 7.2 & 38.0/33 & 9 \\
\hline
\end{tabular} \\
\leftline{\textsuperscript{a} $N_H$ fixed at $2.7\times10^{21}{\rm cm^{-2}}$.}
\leftline{\textsuperscript{b} $0.5-7\,$keV unabsorbed fluxes in units of $10^{-15}{\rm erg\,cm^{-2}s^{-1}}$.}
\end{table}

\section{Discussion and Conclusions}

The shape of the filament is complex and the epoch-to-epoch changes are subtle. We seek to explain these through a combination of variable particle injection at the moving pulsar, particle flow along field lines, particle diffusion across field lines and possible cooling. In practice cooling is likely not important on the scale of the observed filament since standard synchrotron theory gives a cooling time of
\begin{equation}
    \tau \approx 7.6 \times 10^4 E_{\rm keV}^{-1/2} B_{\mu G}^{-3/2} {\rm y}.
\end{equation}
With an observed photon energy of $\sim 2$\,keV and $B_{\mu G} \sim 15$ estimated in \S2.3, we get a cooling time $\tau \approx 930$\,y, so over our four epochs we expect no significant cooling. Since the pulsar moves $\sim 3^\prime$ (twice the size of the Guitar body) in time $\tau$, cooling predicts a fading on this scale. Accordingly, the smaller scale morphology changes must be due to variable injection, advection and diffusion.

In the original \citet{Bandiera2008} picture the $r_c$ relevant for escape was that of the shocked pulsar wind. Since the mean field in that wind increases as $r_0$ decreases, $r_c/r_0$ is essentially constant, and does not control the particle escape; in this picture most bow-shock pulsars should produce filaments and they should do so at all epochs independent of the bow shock size. This does not appear to be the case, since filaments are rare and preferentially associated with pulsars with small $r_0$. Instead we argue that energetic $e^\pm$ are produced via reconnection throughout the shocked pulsar wind and that $r_c$ beyond the contact discontinuity, in the shocked ISM and external medium, controls escaping particle motion. The near-apex external field is modified by the draping effect to have a characteristic curvature radius $r_0$ and thus $r_c/r_0$ in this medium can control which particles move far enough in a gyroradius to encounter different external field orientations, and escape.

The curvature of the filament leading edge implies that the ambient field lines are not completely straight, although the similarity of the edge from epoch to epoch suggests that they are locally approximately parallel. The leading edge is quite sharp. The $e^\pm$ gyroradius $r_c$ in the local field subtends an angle of
\begin{equation}
    \theta =r_c/d \approx 26^{\prime\prime} E_{\rm keV}^{1/2} B_{\mu G}^{-3/2} d_{\rm kpc}^{-1}
\end{equation}
for particles producing a peak photon energy $E_{\rm keV}$. For a leading edge field of $20\mu G$ (Table \ref{tab:specfits}) we get $\theta \approx 0.5^{\prime\prime}$ for the filament. This is comparable to the {\it CXO} resolution (but substantially larger than $r_0$). The filament leading edge stays sharp for the inner and middle zones, spreading primarily in the outer zone. This implies that the cross field diffusion coefficient ahead of the leading edge in the ambient ISM is small. 

With an estimate of the flow speed $v_{e^\pm}$ along the leading edge, one could use the broadening with distance to get an estimate of this forward diffusion coefficient. Noting that this edge is actually the front reached by particles moving rapidly along a set of field lines, we see that the far filament represents earlier injection, onto field lines behind that connect to the pulsar at its current position. Thus the filament front follows an angle $\theta_f \sim v_{\rm PSR}{\rm cos}\Psi/v_{e^\pm}$ behind the ISM field lines, with the field lines themselves at an angle $\Psi\sim 25^\circ$ to the proper motion. If the filament and counter-filament propagation speeds are equal, we can account for $\Psi$ by comparing the PAs of the two sides; these should differ by $2\theta_f$. In practice this measurement is difficult since the counter-filament is short and the section closest to the bow shock suffers PWN-induced distortion. Very roughly, we estimate $\theta_f\lesssim 2^\circ$, and thus $v_{e^\pm} \approx v_{\rm PSR}{\rm cos} \Psi /\theta_f \gtrsim c/13$.

With small cross-field diffusion, we would expect particles confined to their injection field line and the filament would present an approximately uniform band, shifted increasingly farther from the Guitar axis, since particles on field lines to the rear would have more time to propagate away. This band would have brighter ridges marking times (field lines) of enhanced particle injection and a smooth fading on arcmin scales behind the leading edge due to synchrotron cooling. This is not what we see. Instead the emission behind the leading edge is patchy and seems to evolve on times short compared to the cooling times. 

This may be understood if cross-field advection and diffusion increase behind the leading edge. From Figure \ref{fig:spread} we estimate the 2021 surface brightness peak as having position $x_{2021} \approx 12.2^{\prime\prime}$ behind the 2021 leading edge, with bulk motion of $\mu_{\rm ridge} \sim 0.15^{\prime\prime}{\rm y^{-1}}$ and spread of $\sigma(t) \sim [3.0+0.25(t-2000)]^{\prime\prime}$. We can attribute these increased rates to turbulence induced behind the leading edge by the injected particles; this leads to increased scattering and easier cross-field propagation. It then becomes interesting to trace the origin of the ridge that moves through Figure \ref{fig:spread}. With a coordinate increasing normal to and behind the leading edge, we can write the pulsar position at year $t$ as $x_p = \mu_{PSR} {\rm cos}\Psi (2021-t)$.  Similarly $x_{\rm ridge} = x_{2021} + \mu_{ridge} (t-2021)$, with the proper motions in arcsec/y. Finally the propagation time between the pulsar and the middle zone $l \sim 75^{\prime\prime} d \sim 1$\, lt-y away is $t_\parallel \approx l/v_{e^\pm} \sim c/v_{e^\pm}$ years. Thus the date for the enhanced injection of the $e^\pm$ that we see in 2021 as a ridge moving through the filament is  
\begin{equation}
    t_{\rm inj} \approx2021 - x_{\rm r,2021}/(\mu_{\rm PSR} {\rm cos}\Psi+\mu_{\rm ridge}) +t_\parallel .
\end{equation}
From our fit to the ridge evolution we get $t_{\rm inj} \approx 1993_{-16}^{+7} + t_\parallel$, so to identify the moving ridge with particles injected when the pulsar was at the `pinch' in the Guitar head, $7^{\prime\prime}$ behind the apex, in  1985, we would want $\mu_{\rm ridge}$ low in the fit range and $t_\parallel = l/v_{e^\pm}<8 $\, y. Note that $\sigma$ decreases to 0 at $1990_{-10}^{+5}$, so consistent with $\sim 1985$, as well.

It is likely a coincidence that the back-propagation of the ridge brings it nearly parallel with the base of the Guitar head in the 2021 epoch. Although we don't see strong emission at this position in our earlier epochs, those images are shallow, and it is possible that future deep observations will show that this ridge is a permanent feature fixed in space. In that case it would be compatible with the simpler hypothesis that it is the fossil of strong injection at the point the pulsar broke into the head region in $\sim 1940$. More generally the lack of such `fossil' X-ray emission parallel with the Guitar body suggests that when the pulsar was blowing the bubbles corresponding to the body structure, the standoff $r_0$ was large and that little or no $e^\pm$ escape occurred. Thus the Guitar may have had an X-ray filament only since the very compact head region was formed.

Although forward propagation of the pulsar-generated cosmic rays is severely limited at the leading edge, it seems much freer behind, rearranging the injected particles long before they cool. While this means that the filament surface brightness profile is not a simple historical record of injection history, it does offer the opportunity to probe the diffusion of multi-TeV $e^\pm$ through the ISM and, more importantly, their effect, via induced MHD waves, on the local particle propagation. For B2224, injection seems to be effective over the $\sim 20^{\prime\prime}$ region of the Guitar head and neck where the bow shock standoff was evidently small, leading to a wide filament. In contrast, the filament of PSR J2030+4415 stays narrow since the injected period was short and the pulsar covered little distance in this time. 
This picture of variable injection may certainly be tested by finding more example filaments, and connecting them with bow shock properties. Numerical simulations can also be useful in determining whether external field-controlled escape is viable or some other peculiarity of small $r_0$ bow shocks, such as enhanced local turbulence or asymmetric reconnection to the external fields, needs to be invoked.

The effect of injected $e^\pm$ may be especially important in connection with the recently observed TeV halos around nearby young pulsars \citep{2017Sci...358..911A}. Interesting experiments to probe these effects would be enabled by identifying enhanced filament injection events (plausibly via H$\alpha$ bow shock monitoring) followed by a decade of sensitive X-ray images to observe, via synchrotron emission, the TeV $e^\pm$ pulse propagating and spreading. Such a campaign would be expensive in observation time, but would yield a rich harvest of information of cosmic ray and magnetic field dynamics, spread out before the observer in evolving filament images.   

\acknowledgements

We wish to thank the observatory staff who helped in planning the exposures described in this paper, especially Jean Connelly of the CfA for help with {\it CXO} and Alison Vick and Ray Lucas of STScI for help with the ACS/WFC. We also wish to thank Marten van Kerkwijk for advice on the GMOS-N IFU data set.
\smallskip
\vspace{5mm}

MdV and RWR were supported in part by NASA grant G08-19050A, through the Smithsonian Astrophysical Observatory. GGP was supported by NASA grant G08-19050B. OK was supported by NASA  grant GO8-19050C and ADAP grant 80NSSC19K0576. Support for this work was provided by the National Aeronautics and Space Administration through Chandra Award Number GO8-19050 issued by the Chandra X-ray Observatory Center, which is operated by the Smithsonian Astrophysical Observatory for and on behalf of the National Aeronautics Space Administration under contract NAS8-03060.

\vspace{5mm}
\facilities{HST(ACS/WFC), Gemini(GMOS-N), CXO}

\bibliography{B2224}
\bibliographystyle{aasjournal}



\end{document}